\documentclass[12pt]{article}
\newcommand{\be}{\begin{equation}}
\newcommand{\ee}{\end{equation}}
\newcommand{\bea}{\begin{eqnarray}}
\newcommand{\eea}{\end{eqnarray}}
\newcommand{\sptwo}{1.4}
\newcommand{\doublespace}{\edef\baselinestretch{\sptwo}\Large\normalsize}
\newcommand{\newsection}[1]{
\section{#1}
\setcounter{equation}{0}}

\newcounter{newapp}
\begin{document}
\vspace*{0.2in}
\begin{center}
{\large\bf Oscillating p-Branes}
\end{center}
\vspace{0.2in}
\begin{center}
{T.E. Clark}\footnote{e-mail address: clark@physics.purdue.edu}$~^a~,~${S.T. Love}\footnote{e-mail address: loves@physics.purdue.edu}$~^{a}~,~${Muneto Nitta}\footnote{e-mail address: nitta@phys-h.keio.ac.jp}$~^b~,~${T. ter Veldhuis}\footnote{e-mail address: terveldhuis@macalester.edu}$~^{a,c}~,~${C. Xiong}\footnote{e-mail address: xiong@purdue.edu}$~^a$\\
\end{center}
\begin{center}
{{\bf a.}~\it Department of Physics,\\
 Purdue University,\\
 West Lafayette, IN 47907-2036, U.S.A.}\\
\end{center}
\begin{center}
{{\bf b.}~\it Department of Physics,\\
 Keio University,\\
 Hiyoshi, Yokohoma, Kanagawa, 223-8521, Japan}\\
\end{center}
\begin{center}
{{\bf c.}~\it Department of Physics \& Astronomy,\\
 Macalester College,\\
 Saint Paul, MN 55105-1899, U.S.A.}
\end{center}
~\\
~\\
\begin{center}
{\bf Abstract}
\end{center}
Coset methods are used to construct the action describing the dynamics associated with the spontaneous breaking of the Poincar\'e symmetries of $D$ dimensional space-time due to the embedding of a $p$-brane with codimension $N=D-p-1$.  The resulting world volume action is an $ISO(1,p+N)$ invariant generalization of the Nambu-Goto action in $d=p+1$ dimensional space-time.  Analogous results are obtained for an $AdS$ $p$-brane with codimension $N$ embedded in $D$ dimensional $AdS$ space, yielding an $SO(2,p+N)$ invariant version of the Nambu-Goto action in $d=p+1$ dimensional space-time. Attention is focused on a supersymmetric extension of the $D=6$ Minkowski space case with an embedded $p=3$ brane; a particular realization of which is provided by a non-BPS vortex. Here both the Nambu-Goto-Akulov-Volkov action and its dual tensor form are presented.

~\\
\pagebreak
\doublespace

\newsection{\large Brane Oscillations In Minkowski Space}

The nonlinear realization of spontaneously broken space-time symmetries provides a systematic method of obtaining dynamical properties of a lower dimensional probe brane whose presence is responsible for the symmetry breakdown.  In the case that the bulk is $D$ dimensional Minkowski space, the $p$-brane defect breaks the isometry group from $ISO(1,D-1)$ down to that of the $d=(1+p)$ dimensional world volume and its $N=(D-d)$ dimensional complement, $ISO(1,p)\times SO(N)$.  The long wavelength modes of the $p$-brane are described by the Nambu-Goldstone
bosons associated with the collective coordinate translations transverse to
the brane into the $N$ dimensional covolume.  Indeed, the Nambu-Goto action governing the zero mode fields' dynamics is readily obtained in a model independent way by nonlinearly
realizing the broken symmetries on the Nambu-Goldstone fields \cite{Coleman:sm,Volkov73}.  The symmetry generators of $ISO(1, D-1)$ are the $D$ dimensional translations $P^{\cal M}$ and Lorentz transformations $M^{\cal MN}$ where ${\cal M,N} = 0,1, \ldots ,p+N$.  For the case of a $p$-brane with codimension $N$, the $d$ dimensional world volume Poincar\'e generators, $P^{\mu}$ for space-time translations and $M^{\mu\nu}$ for Lorentz rotations, with $\mu , \nu = 0,1,\ldots , p$, form an unbroken subgroup $ISO(1, p)$ of the $D$ dimensional Poincar\'e group $ISO(1, D-1)$.  In addition, the rotations in the $N$ dimensional covolume with generators $T^{ij}=M^{p+i, p+j}$, where $i,j =1, 2, \ldots N$, form an unbroken subgroup $SO(N)$.  The broken symmetry charges are the generators of translations transverse to the brane, denoted by $Z_i = -P^{p+i}$, and the generators of the broken bulk Lorentz transformations, denoted by $K^\mu_i = 2 M^{p+i, \mu}$.  The $d$ dimensional Lorentz scalars $Z_i$ are in the fundamental (vector) representation of the unbroken $SO(N)$ subgroup.  Likewise, the $d$ dimensional Lorentz vector $K^\mu_i$ is also a $SO(N)$ vector.  The explicit form of the charge algebra is given in \cite{Clark:2006pd} and can be found as the Minkowski limit of the $AdS$ algebra of equations (\ref{SO2DAlgebra})-(\ref{Algebra3}).  

The $D$ dimensional Poincar\'e group of transformations can be realized by group elements acting on the $ISO(1, D-1)/SO(1, p)\times SO(N)$ coset element $\Omega$ which is constructed from the $P^\mu ,~Z_i ,~K^\mu_i$ charges as
\be
\Omega (x) \equiv e^{i x^\mu P_\mu } e^{i \phi^i(x) Z_i} e^{iv^\mu_i(x) K_\mu^i} .
\ee
Here $x^\mu$ are the $d$ dimensional world volume space-time coordinates of the $p$-brane in the
static gauge, while $\phi^i (x)$ and $v^\mu_i (x)$ are the collective
coordinate Nambu-Goldstone bosons associated with the broken $D$ dimensional Poincar\'e
symmetries and correspond to the massless excitation modes of the brane.  Left multiplication of the coset elements $\Omega$ by an $ISO(1,D-1)$ group element $g$, which is specified by global transformation parameters $\epsilon^\mu , z^i, b_i^\mu , \lambda^{\mu\nu}, \theta^{ij}$, so that 
\be
g(x) = e^{i\epsilon^\mu P_\mu} e^{iz^i Z_i} e^{ib^\mu_i K^i_\mu} e^{\frac{i}{2} \lambda^{\mu\nu} M_{\mu\nu}}e^{\frac{i}{2} \theta^{ij} T_{ij}},
\ee
results in transformations of the space-time coordinates and the Nambu-Goldstone fields according to the general form \cite{Coleman:sm}
\be
g(x)\Omega(x) = \Omega^\prime(x^\prime) h(x) .
\label{leftmult}
\ee
The transformed coset element, $\Omega^\prime (x^\prime)$,  is a function of the transformed world volume coordinates and the total variations of the fields
\be
\Omega^\prime (x^\prime) = e^{ix^{\prime\mu}  P_\mu} e^{i\phi^{\prime i}(x^\prime)Z_i} e^{iv^{\prime\mu}_i (x^\prime) K^i_\mu},
\ee 
while $h(x)$ is a field dependent element of the stability group $SO(1,p)\times SO(N)$:
\be
h(x)= e^{\frac{i}{2} \alpha_{\mu\nu}(x) M^{\mu\nu}}e^{\frac{i}{2} \beta_{ij}(x) T^{ij}} .
\label{hele}
\ee

Exploiting the algebra of the $ISO(1,D-1)$ charges, along with extensive use of the Baker-Campbell-Hausdorff formulae, the $ISO(1, D-1)$ transformations are obtained for infinitesimal transformation parameters as
\bea
x^{\prime\mu}  &=&  (\eta^{\mu\nu} -\lambda^{\mu\nu})x_\nu +\epsilon^\mu  +2b^\mu_i  \phi^i (x)\cr
 & & \cr
\phi^{\prime}_i (x^\prime) &=& (\delta_{ij} +\theta_{ij})\phi^j (x) + z_i + 2b^\mu_i  x_\mu \cr
 & & \cr
v^{\prime \mu}_i (x^\prime) &=& (\eta^{\mu\nu} - \lambda^{\mu\nu})(\delta_{ij} + \theta_{ij})v_{\nu j} (x) -b_j^\nu  M_{\nu k}^{j \rho}(x) \left(\coth{\sqrt{M (x)}}\right)_{\rho i}^{k \mu} \cr
 & & \cr
\alpha^{\mu\nu}(x) &=& \lambda^{\mu\nu} -4b_j^\lambda  [M^{-1/2}(x)]^{j \rho}_{\lambda k}  \cr
 & &\left\{ \left(\tanh{\sqrt{M(x)/4}}\right)^{k \mu}_{\rho i} v_i^\nu (x) -\left(\tanh{\sqrt{M(x)/4}}\right)^{k \nu}_{\rho i} v_i^\mu (x)\right\} \cr
\beta_{ij}(x) &=& \theta^{ij} -4b_l^\lambda  [M^{-1/2}(x)]^{l \rho}_{\lambda k} \cr
 & & \left\{ \left(\tanh{\sqrt{M(x)/4}}\right)^{k \mu}_{\rho j} v_{\mu i} (x) -\left(\tanh{\sqrt{M(x)/4}}\right)^{k \mu}_{\rho i} v_{\mu j} (x)\right\} ,
\label{variations}
\eea
where the matrix $M$ is defined as
\be
M_{ij}^{\mu\nu} =  4\left( \eta^{\mu\nu} U_{ij} - 2 v^\mu_j v^\nu_i + W^{\mu\nu}\delta_{ij}\right) ,
\label{Mmatrix}
\ee
with 
\bea
U_{ij} &=& v^\mu_i \eta_{\mu\nu} v^\nu_j \cr
W^{\mu\nu} &=& v^\mu_i \delta^{ij} v^\nu_j
\label{UW}
\eea
and $\eta_{\mu\nu}$ is the $d$ dimensional Minkowski space metric tensor with signature $(+1,-1,\ldots,-1)$.  In the above, the space-time indices are raised, lowered and contracted using $\eta_{\mu\nu}$ while the Kronecker $\delta_{ij}$ is similarly used for the $SO(N)$ indices.  Both Nambu-Goldstone fields $\phi^i$ and $v^\mu_i$ transform inhomogeneously under the broken translations $Z^i$ and broken local Lorentz transformations $K^\mu_i$.

The nonlinearly realized $ISO(1,D-1)$ transformations induce a coordinate and field dependent general coordinate transformation of the world volume space-time coordinates.  From the $x^\mu$ coordinate transformation given above, the general coordinate Einstein transformation for the world volume space-time coordinate differentials is given by
\be
dx^{\prime \mu} = dx^\nu {G}_\nu^{~\mu} (x),
\label{dxprime}
\ee
with ${G}_\nu^{~\mu}(x) = \partial x^{\prime \mu}/\partial x^\nu$.  The $ISO(1,D-1)$ invariant interval can be formed using the metric tensor ${g}_{\mu\nu}(x)$ so that $ds^2 = dx^\mu {g}_{\mu\nu}(x) dx^\nu = ds^{\prime 2} = dx^{\prime \mu} {g}^\prime_{\mu\nu}(x^\prime) dx^{\prime \nu}$ where the metric tensor transforms as 
\be
{g}^\prime_{\mu\nu} (x^\prime) = {G}_\mu^{-1\rho}(x) {g}_{\rho\sigma}(x) {G}_\nu^{-1\sigma}(x) .
\label{gprime}
\ee

The form of the vielbein (and hence the metric tensor) as well as the $ISO(1,D-1)$
covariant derivatives of the Nambu-Goldstone boson fields and the spin and $SO(N)$ connections can be extracted from the Maurer-Cartan one-form, $\Omega^{-1}d\Omega$, which can be expanded in terms of the generators as 
\be
\omega (x) =\Omega^{-1} d\Omega = i\left( \omega^m P_m + \omega_{Zi} Z^i +\omega^m_{Ki} K_{im} +\frac{1}{2}\omega_M^{mn} M_{mn}+\frac{1}{2}\omega_{T}^{ij} T_{ij}\right).
\label{covMC1forms}
\ee
Here Latin indices $m,n = 0,1,\ldots ,p$, are used to distinguish tangent space induced local Lorentz transformation properties from world volume Einstein transformation properties which are denoted using Greek indices. In what follows, Latin indices are raised and lowered with use of the Minkowski metric tensors, $\eta^{mn}$ and $\eta_{mn}$, while Greek indices are raised and lowered with use of the induced space-time metric tensors, $g^{\mu\nu}$ and $g_{\mu\nu}$.  

Using the transformation property of the coset element, equation (\ref{leftmult}), the Maurer-Cartan one-form transforms according to 
\be
\omega^\prime(x^\prime) = h(x)\omega (x) h^{-1}(x) +h(x)dh^{-1}(x),
\ee
with $h(x)= e^{\frac{i}{2}[\alpha^{mn}(x) M_{mn} +\beta^{ij}(x) T_{ij}]}$ as given in Eqs. (\ref{hele}) and (\ref{variations}).  Expanding in terms of the $D$ dimensional Poincar\'e charges, the individual one-forms transform according to their local Lorentz and $SO(N)$ nature so that
\bea
\omega^{\prime m}(x^\prime) &=&  \omega^n (x)\Lambda^{~m}_{n}(\alpha (x)) \cr
\omega_{Zi}^\prime (x^\prime)&=& R_{ij} (\beta(x))\omega_{Zj} (x)\cr
\omega^{\prime m}_{Ki} (x^\prime)&=&  R_{ij} (\beta (x))\omega^n_{Kj} (x)\Lambda^{~m}_{n}(\alpha (x))\cr
\omega^{\prime mn}_M (x^\prime)&=& \omega^{rs}_M (x)\Lambda^{~m}_{r}(\alpha (x))\Lambda^{~n}_{s}(\alpha (x))-d\alpha^{mn}(x) \cr
\omega^{\prime}_{Tij} (x^\prime)&=& R_{ik} (\beta (x)) R_{jl} (\beta (x)) \omega_{Tkl}(x)-d\beta_{ij}(x) .
\label{oneformvari}
\eea
For infinitesimal transformations, the local Lorentz transformations are $\Lambda^{~m}_{n}(\alpha (x)) = \delta^{~m}_{n} + \alpha^{~m}_{n}(x)$ and the local $SO(N)$ transformations are $R_{ij} (\beta (x))= \delta_{ij} +\beta_{ij} (x)$.   

Using the Feynman formula for the variation of an exponential operator in conjunction with the Baker-Campell-Hausdorff formulae, the individual world volume one-forms appearing in the above decomposition of the covariant Maurer-Cartan one-form are secured as 
\bea
\omega^m &=& dx^\mu e_\mu^{~m}= dx^\mu \left\{\delta_\mu^{~m} +v_{\mu j} \left(U^{-1/2}(\cosh{\sqrt{4U}}-1)U^{-1/2}\right)_{ji} v_i^m \right.  \cr
 & & \left.\qquad\qquad - \partial_\mu \phi_j \left(U^{-1/2}\sinh{\sqrt{4U}}\right)_{ji} v_i^m \right\}\cr
\omega_{Zi} &=& dx^\mu \omega_{Zi\mu} \cr
 &=& dx^\mu \left(\cosh{\sqrt{4U}}\right)_{ij}  \left\{\partial_\mu\phi_j  -\left(U^{-1/2}\tanh{\sqrt{4U}}\right)_{jk}v_{k\mu}  \right\}\cr
\omega^m_{Ki} &=& dx^\mu \omega_{Ki\mu}^m \cr
 &=&  dx^\mu (\partial_\mu v^n_j) \left(\sinh{\sqrt{M}}M^{-1/2}\right)^{jm}_{ni} \cr
\omega_M^{mn} &=& dx^\mu \omega_{M\mu}^{mn}\cr
 &=& -4dx^\mu (\partial_\mu v^r_i) \left\{ \left[\left(\cosh{\sqrt{M}}-1 \right)M^{-1}  \right]^{im}_{rj} v_j^n  -\left[\left(\cosh{\sqrt{M}}-1 \right)M^{-1}  \right]^{in}_{rj} v_j^m   \right\}\cr
\omega_{Tij} &=& dx^\mu \omega_{Tij\mu}\cr
 &=& -4dx^\mu (\partial_\mu v^m_l) \left\{ \left[\left(\cosh{\sqrt{M}}-1 \right)M^{-1}  \right]^{ln}_{mj} v_{ni}  -\left[\left(\cosh{\sqrt{M}}-1 \right)M^{-1}  \right]^{ln}_{mi} v_{nj}   \right\}.\cr
 & & 
\label{MCOne-form}
\eea
The covariant coordinate differential $\omega^m$ is related to the world volume coordinate differential $dx^\mu$ by the vielbein, $e_\mu^{~m}$, as $\omega^m = dx^\mu e_\mu^{~m}$.

The vielbein transforms as $e_\mu^{\prime m} (x^\prime) = G_{\mu}^{-1\nu}(x)
e_\nu^{~n}(x)\Lambda^{~m}_{n} (\alpha (x))$.  Since the Jacobian of the $x^\mu \rightarrow x^{\prime \mu}$ transformation is simply
\bea
d^dx^\prime &=& d^d x ~\det{{G}},  
\eea
it follows that $d^dx^\prime ~\det{e^\prime} (x^\prime) = d^dx ~\det{e} (x)$ since $\det{\Lambda} = 1$.
Thus an $ISO(1, D-1)$ invariant action \cite{Gomis:2006xw} for the Nambu-Goto fields is constructed as
\be
\Gamma = -\sigma\int d^d x \det{e(x)} ,
\label{action1}
\ee
with $\sigma$ the brane tension.  

As seen from the expression for the vielbein, the action does not contain any derivatives of the vector field $v_\mu^i$.  Its equation of motion implies the covariant constraint of the inverse Higgs mechanism \cite{Ivanov:1975zq}, $\omega_{Zi}=0$.  This allows $v^i_\mu$ to be expressed in terms of $\partial_\mu \phi^i$
\be
\partial_\mu \phi^i = \left( U^{-1/2} \tanh{\sqrt{4U}}\right)_{ij} v_\mu^j.
\ee
Applying this to the vielbein gives
\be
e_\mu^{~m} = \delta_\mu^{~m} + \partial_\mu \phi^i G_{ij} \partial^m \phi^j ,
\label{e1}
\ee
where
\be
G_{ij}\equiv H^{-1/2}_{ik}\left[\sqrt{(1-H)}-1\right]_{kl} H^{-1/2}_{lj} 
\label{G}
\ee
is an $SO(N)$ symmetric matrix with 
\be
H_{ij}\equiv \partial_\mu \phi_i \partial^\mu \phi_j.
\label{H}
\ee
This in turn yields the Nambu-Goto form of the $ISO(1,D-1)$ invariant action for the Nambu-Goldstone fields $\phi^i$ as
\bea
\Gamma &=& -\sigma \int d^d x \det{e} \cr
 &=&-\sigma \int d^d x \sqrt{\det{(\delta^{ij}-\partial_\mu \phi^i \partial^\mu \phi^j)}}.
\label{action2}
\eea

The form of this action can also be secured by considering the $ISO(1,D-1)$ invariant interval in $D$ dimensional Minkowski space with coordinates $X^{\cal M}=(x^\mu , \phi^i (x))$ which reads
\be
ds^2 = dX^{\cal M} \eta_{\cal MN} dX^{\cal N} = dx^\mu \eta_{\mu\nu} dx^\nu -d\phi^i (x) \delta_{ij}d\phi^j (x),
\ee
where the static gauge coordinate oscillations of the $p$-brane into the covolume are identified with the $N$ Nambu-Goldstone fields $\phi^i (x)$.  This induces a metric $g_{\mu\nu}$ on the world volume as
\be
ds^2 = dx^\mu \left[ \eta_{\mu\nu} - \partial_\mu \phi^i \delta_{ij}\partial_\nu \phi^j\right]dx^\nu =dx^\mu g_{\mu\nu}dx^\nu.
\label{metricind}
\ee
Consequently, the $ISO(1,D-1)$ invariant action is again given as in equation (\ref{action2}) takes the form
\be
\Gamma = -\sigma \int d^d x \sqrt{(-1)^p\det{g..}}=
-\sigma \int d^d x \sqrt{\det{(\delta^{ij}-\partial_\mu \phi^i \partial^\mu \phi^j)}}.
\ee
The induced Nambu-Goto vielbein $e_\mu^{~m}$ is just the one given in (\ref{e1}) which produces the metric in equation (\ref{metricind}), $g_{\mu\nu}= e_\mu^{~m} \eta_{mn} e_\nu^{~n}$.

\newsection{Probe Branes In $AdS$ Space}
Embedding a $AdS_d$ probe brane in $AdS_D$ space breaks the $SO(2,D-1)$ symmetry of the $AdS_D$ bulk space down to the $SO(2, d-1)$ isometry group of the $AdS_d$ world volume and its complement: $SO(2,p)\times SO(N)$.  The $SO(2,D-1)$ isometries of the $AdS_D$ bulk space are generated by the $D$ dimensional (pseudo-) translation operators, denoted by $P^{\cal M}$, and the Lorentz transformations, $M^{\cal MN}$, ${\cal M,N} = 0,1, \ldots ,p+N$, collectively satisfying the algebra \cite{Clark:2005ht}
\bea
\left[M^{\cal MN} , M^{\cal RS} \right] &=& -i \left( \eta^{\cal MR} M^{\cal NS} -\eta^{\cal MS} M^{\cal NR} +\eta^{\cal NS} M^{\cal MR} -\eta^{\cal NR} M^{\cal MS} \right) \cr
\left[M^{\cal MN} , P^{\cal L} \right] &=& i \left( P^{\cal M} \eta^{\cal NL} - P^{\cal N} \eta^{\cal ML} \right) \cr
\left[P^{\cal M} , P^{\cal N} \right] &=& -i m^2 M^{\cal MN} .
\label{SO2DAlgebra}
\eea
Here $m^2>0$ is a constant inverse squared length scale characterizing the $AdS$ bulk space as well as the $AdS$ world volume.  The Minkowski space case with symmetry breaking pattern $ISO(1,D-1)\rightarrow ISO(1,d-1)$ is secured as  the  $m^2=0$ limit. For an $AdS$ $p$-brane with codimension $N$, the $d$ dimensional world volume symmetry generators, $P^{\mu}$ for (pseudo-) translations and $M^{\mu\nu}$ for Lorentz rotations, with $\mu , \nu = 0,1,\ldots , p$, form an unbroken subgroup $SO(2, p)$ of the $D$ dimensional $AdS_D$ symmetry group $SO(2, D-1)$. The specific case with one codimension, $N=1$, was discussed in \cite{Clark:2005ht}. The $N$ dimensional covolume rotation generators, $T^{ij}=M^{p+i, p+j}$, where $i,j =1, 2, \ldots N$, form an unbroken subgroup $SO(N)$.  The broken symmetry charges are the generators of (pseudo-) translations transverse to the brane, denoted by $Z_i = -P^{p+i}$, and the generators of the broken bulk Lorentz transformations, denoted by $K^\mu_i = 2 M^{p+i, \mu}$. Both the $d$ dimensional Lorentz scalars $Z_i$ and the $d$ dimensional Lorentz vector $K^\mu_i$ transform according to the fundamental (vector) representation of the unbroken $SO(N)$ subgroup.  The  unbroken generators of the $SO(2,D-1)$ algebra, (c.f. equation (\ref{SO2DAlgebra})), satsify 
\bea
\left[M^{\mu\nu} , M^{\rho\sigma} \right] &=& -i \left( \eta^{\mu\rho} M^{\nu\sigma} -\eta^{\mu\sigma} M^{\nu\rho} +\eta^{\nu\sigma} M^{\mu\rho} -\eta^{\nu\rho} M^{\mu\sigma} \right) \cr
\left[M^{\mu\nu} , P^{\lambda} \right] &=& i \left( P^\mu \eta^{\nu\lambda} - P^\nu \eta^{\mu\lambda} \right) \cr
\left[P^\mu , P^\nu \right] &=& -i m^2 M^{\mu\nu} \cr
\left[T^{ij} , T^{kl} \right] &=& i \left( \delta^{ik} T^{jl} -\delta^{il} T^{jk} +\delta^{jl} T^{ik} -\delta^{jk} T^{il} \right) \cr
\left[P^\mu , T^{ij} \right] &=& 0 = \left[M^{\mu\nu} , T^{ij} \right] ,
\eea
while the broken generators carry representations of the unbroken symmetries and rotate the unbroken into broken charges as
\bea
\begin{array}{ll}
\left[M^{\mu\nu} , Z_{i} \right] =0 &
\left[M^{\mu\nu} , K^{\lambda}_i \right] = i \left( K^\mu_i \eta^{\nu\lambda} - K^\nu_i \eta^{\mu\lambda} \right)\\
\left[T^{ij} , Z_{k} \right] = -i \left( Z_i \delta^{jk} - Z_j \delta^{ik} \right) &
 \left[T^{ij} , K^{\mu{k}} \right] = -i \left( K^{\mu i} \delta^{jk} - K^{\mu j} \delta^{ik} \right) \\
\left[P^\mu , Z_{i} \right] = -\frac{i}{2}m^2 K^\mu_i & 
\left[K^\mu_i , P^\nu  \right] = -2i \eta^{\mu\nu} Z_i .\\
\end{array} & & 
\eea
Finally, the commutation relations between broken generators take the form
\bea
\left[Z_i , Z_j \right]  & =  & -im^2 T^{ij}  \nonumber \\
\left[K^\mu_i , K^\nu_j \right] & = & 4i \left( \delta_{ij} M^{\mu\nu}-\eta^{\mu\nu} T^{ij} \right) \nonumber \\
\left[K^\mu_i , Z_j \right] & = & -2i \delta_{ij}P^\mu. 
\label{Algebra3}
\eea
These results are instrumental in deriving the $SO(2,D-1)$ invariant action describing the motion of the $AdS_d$ probe brane in $AdS_D$ space which will be obtained using three different constructions. In section (\ref{section1}) the coset method is used, while in (\ref{section2}) an embedding method is employed. Finally in section  (\ref{section3}) the action is obtained through a hybrid of these two methods.

\subsection{Coset method \label{section1}}
The invariant action of the Nambu-Goldstone fields along with their $SO(2,D-1)$ transformations and those of the coordinates can be obtained using coset methods. Parametrizing a coset element as 
\be
\Omega (x) \equiv e^{i x^\mu P_\mu } e^{i \phi^i(x) Z_i} e^{iv^\mu_i(x) K_\mu^i} 
\ee
and expanding the Maurer-Cartan one-form as in equation (\ref{covMC1forms}), the veilbein $\omega^m = dx^\mu e_\mu^{~m}$ and the $SO(2, D-1)$ covariant derivative of $\phi^i$, $\omega_{Z}^i =dx^\mu \nabla_\mu \phi^i$, are extracted as
\bea
e_\mu^{~m} &=& \cosh{\sqrt{m^2\phi^2}}~\bar{e}_\mu^{~n} \left\{ \delta_{n}^{~m} +v_{in} \left[ U^{-1/2} \left( \cosh{\sqrt{4U}}-1\right)U^{-1/2} \right]_{ij} v^m_j \right. \cr
 & & \left. - {\cal D}_m\phi^i \left( U^{-1/2} \sinh{\sqrt{4U}}\right)_{ij} v^m_j\right\} \cr
 & & \cr
\nabla_\mu \phi^i &=& -\cosh{\sqrt{m^2\phi^2}}~\bar{e}_\mu^{~n}\left\{ \left( U^{-1/2} \sinh{\sqrt{4U}}\right)_{ij} v_{nj} +{\cal D}_n\phi^j (\cosh{\sqrt{4U}})_{ji} \right\}.\cr
 & & \label{adsmcforms}
\eea
Here $\bar{e}_\mu^{~m}$ is the $AdS_d$ background vielbein \cite{Clark:2005ht} defined as
\be
\bar{e}_\mu^{~m}(x) = \frac{\sin{\sqrt{m^2x^2}}}{\sqrt{m^2x^2}} P_{T\mu}^{~~~m}(x)  +P_{L\mu}^{~~~m}(x) , \label{AdS_dvielbein}
\ee
and the transverse and longitudinal projectors for $x^\mu$ are
\bea
P_{T\mu\nu}(x)&=& \eta_{\mu\nu} -\frac{x_\mu x_\nu}{x^2} \cr
P_{L\mu\nu}(x)&=&\frac{x_\mu x_\nu}{x^2} . \label{xprojectors}
\eea
The partially covariant derivative ${\cal D}_m \phi^i$ is defined as
\be
{\cal D}_m \phi^i = e^{-A(\phi)} \bar{e}_j^{~i}\bar{e}^{-1\mu}_m \partial_\mu \phi^j , \label{PartCovDer}
\ee
where $A(\phi)= \ln{\cosh{\sqrt{m^2\phi^2}}}$ is the warp factor with $\phi^2 = \phi^i \phi^i$.
In addition, $\bar{e}_i^{~j}$, which is identified in section (\ref{section3}) as the Euclidean $AdS_N$ background covolume vielbein, takes the form
\be
\bar{e}_{ij} = \frac{\sinh{\sqrt{m^2\phi^2}}}{\sqrt{m^2\phi^2}} P_{Tij}(\phi) + P_{Lij} (\phi) , \label{AdS_Nvielbein}
\ee
where the transverse and longitudinal projectors for $\phi^i$ are
\bea
P_{T ij}(\phi)&=& \delta_{ij} -\frac{\phi_i\phi_j}{\phi^2} \cr
P_{L ij}(\phi)&=&\frac{\phi_i\phi_j}{\phi^2} . \label{phiprojectors}
\eea
An $SO(2, D-1)$ invariant action is then constructed as
\be
\Gamma = -\sigma \int d^dx \det{e}~.
\ee
Since this action is independent of $v^i_m$ space-time derivatives, the $v^i_m$ field equation takes the form of the constraint $\nabla\phi^i=0$ and allows the non-propagating $v^i_m$ degrees of freedom to be eliminated  in terms of the partially covariant derivative of $\phi^i$ as
\be
{\cal D}_m \phi^i = \left(U^{-1/2}\tanh{\sqrt{4U}} \right)_{ij} v_m^j .
\label{inverseH}
\ee
Using equation (\ref{inverseH}) in the expression for the vielbein $e_\mu^{~m}$, c.f. equation (\ref{adsmcforms}), then gives the factorized form of the vielbein
\bea
e_{\mu}^{~m}(x) & = & e^{A(\phi)}\bar{e}_\mu^{~n}(x) N_n^{~m}(x) \cr
 &=& \cosh{\sqrt{m^2\phi^2}}~\bar{e}_\mu^{~n}~ \left[\delta_n^{~m}  +{\cal D}_n \phi^i(x) G_{ij}{\cal D}^m \phi^j(x)\right] ,
\label{e4}
\eea
where the Nambu-Goto vielbein, $N_n^{~m}$, is given by
\bea
N_n^{~m}(x) &=& \delta_n^{~m}  +{\cal D}_n \phi^i(x) G_{ij}{\cal D}^m \phi^j(x),
\eea
and, as before, the warped $SO(N)$ matrix $G_{ij}$ is 
\be
G_{ij} = H^{-1/2}_{ik}\left[ \sqrt{\left( 1-H \right)} -1\right]_{kl} (H^{-1/2})_{lj},
\ee
but now with $H_{ij}$  defined in terms of the  partially covariant derivatives as
\be
H_{ij} = {\cal D}_r \phi^i(x) \eta^{rs} {\cal D}_s \phi^j(x) .
\ee
Hence the $SO(2,D-1)$ invariant action for the $AdS_d$ $p$-brane takes the form
\be
\Gamma =-\sigma \int d^d x \det{\bar{e}}~\cosh^d{\sqrt{(m^2 \phi^2)}}\sqrt{\det{\left[\delta^{ij} - {\cal D}_m \phi^i \eta^{mn} {\cal D}_n \phi^j  \right]}} .\label{action1a}
\ee

\subsection{Embedding method \label{section2}}
Alternatively, the embedding of the probe brane in the $AdS_D$ space can be implemented by picking specific coordinates with which to describe the $AdS_d$ and covolume subspaces.  The $AdS_D$ space can be simply described as the $SO(2,D-1)$ invariant hyperboloidal hypersurface 
\be
\frac{1}{m^2} = X_0^2 -X_1^2 -X_2^2-\cdots -X_{D-1}^2 +X_{D}^2 = X^{\cal M} \hat\eta_{\cal MN} X^{\cal N} ,
\label{AdSD}
\ee
embedded in a $(D+1)$-dimensional pseudo-Euclidean space defined with invariant interval
\be
ds^2 = dX^{\cal M} \hat\eta_{\cal MN} dX^{\cal N} 
\label{interval}
\ee
characterized by the metric tensor $\hat\eta_{\cal MN}$ of signature $( +1, -1,-1,\ldots,-1,+1 )$, where here ${\cal M, N} = 0,1,\ldots,D-1, D$ and $X^{\cal M}$ are the pseudo-Euclidean space homogeneous coordinates.  

$AdS_{D}$ space containing an $AdS_d$ $p$-brane, which is a $d$ dimensional world volume with an $AdS$ metric embedded as a hypersurface at the covolume coordinates $X^{p+i} =0$ for $i=1,2,\ldots ,N$, can be described by the coordinates $X^{\cal M} = (X^\mu, X^{p+i} , X^{D})$ with isotropic $AdS_d$ coordinates $X^\mu$ and covolume direction cosine coordinates $X^{p+i}$
\bea
X^\mu &=& a(x^2) x^\mu \cosh{\sqrt{(m^2 \phi^2)}}~:~ \mu=0,1,...,p \cr
X^{p+i} &=& \frac{1}{m} \frac{\phi^i}{\sqrt{\phi^2}}\sinh{\sqrt{(m^2 \phi^2)}}~:~i=1,2,\ldots ,N \cr
X^{D} &=& \frac{1}{m} b(x^2) \cosh{\sqrt{(m^2 \phi^2)}} .
\label{coordinates}
\eea
Here $x^\mu$ are the intrinsic coordinates of the $AdS_d$ world volume and the $\phi^i$ are the covolume coordinates. To satisfy the equation (\ref{AdSD}) of the $AdS_{D}$ hyperbola,  $a(x^2)$ and $b(x^2)$ are related as
\be
1 = m^2 x^2 a^2(x^2) + b^2(x^2) .
\label{abR}
\ee
Hence the $SO(2,D-1)$ invariant interval, equation (\ref{interval}), becomes
\be
ds^2 = e^{2A(r)} d\bar{s}^2 -dr^2 ,
\label{KR}
\ee
where the warp factor is $A(\phi)= \ln{\cosh{\sqrt{(m^2\phi^2)}}}$ and $d\bar{s}^2 = dx^\mu \bar{g}_{\mu\nu}(x) dx^\nu$ is the $AdS_d$ invariant interval with $\bar{g}_{\mu\nu}$ the $AdS_d$ metric tensor to be given below. The interval on the covolume is found to be
\be
dr^2 = d\phi^i \bar{g}_{ij}d\phi^j ,
\ee
where the metric $\bar{g}_{ij}$ is found to be
\be
\bar{g}_{ij} = \bar{e}_i^{~k} \delta_{kl}\bar{e}_j^{~l} = \frac{\sinh^2{\sqrt{(m^2\phi^2)}}}{m^2\phi^2} P_{Tij} (\phi) + P_{Lij} (\phi) ,
\ee
with the transverse and longitudinal projectors for $\phi_i$ defined as in equation (\ref{phiprojectors}).

The $AdS_d$ subspace has the isotropic coordinates $x^\mu$ of an $SO(2,p)$ invariant hyperboloid, $\frac{1}{m^2} = X_0^2 -X_1^2 -\cdots -X_{p}^2 +X_{D}^2$, embedded at $\phi^i =0=X^{p+i}$.  This subsurface maintains the coordinate relation equation (\ref{abR}).  This in turn leads to a form for the $AdS_d$ metric tensor given by
\bea
\bar{g}_{\mu\nu} (x) &=& a^2(x^2) P_{T\mu\nu}(x) + \left[ \left(a(x^2) +2x^2 \frac{da(x^2)}{dx^2} \right)^2 + 4\frac{x^2}{m^2} \left(\frac{db(x^2)}{dx^2}\right)^2\right] P_{L\mu\nu}(x) \cr
&=&a^2(x^2)P_{T\mu\nu}(x)+\frac{\left(a(x^2)+2x^2\frac{da(x^2)}{dx^2}\right)^2}{\left(1-m^2x^2a^2(x^2)\right)}P_{L\mu\nu}.
\label{gbar}
\eea
With the specific choice of parameters
\bea
a(x^2) &=& \frac{\sin{\sqrt{m^2x^2}}}{\sqrt{m^2x^2}} \cr
b(x^2) &=& \cos{\sqrt{m^2x^2}} ,
\eea
the $AdS_d$ background vielbein, equation (\ref{AdS_dvielbein}), and metric are obtained in the same coordinate system as implied by the coset method in section (\ref{section1}). 
The induced metric on the probe brane's world volume, $g_{\mu\nu}$, with $SO(2,D-1)$ isometries follows from
\be
ds^2 =dx^\mu g_{\mu\nu}dx^\nu = dx^\mu \left[ e^{2A(\phi)}\bar{g}_{\mu\nu} - \partial_\mu \phi^i \bar{g}_{ij}\partial_\nu \phi^j\right]dx^\nu ,
\ee
and the $SO(2,D-1)$ invariant action is therefore given by
\be
\Gamma = -\sigma \int d^d x \sqrt{(-1)^p\det{g}} ~~, \label{action2a}
\ee
in agreement with equation (\ref{action1a}).
\subsection{Hybrid method \label{section3}}
The $SO(2,D-1)$ isometry group of $AdS_D$ space contains as subgroups $SO(2,p)$, generated by $M^{\mu\nu}$ and $P^\mu$, and $SO(1,N)$, generated by $T^{ij}$ and $Z^i$, which are the isometry groups of an $AdS_d$ subspace and its Euclidean $AdS_N$ covolume, repectively. Below, the coset method is used to obtain the background metric and invariant interval in each of these two subspaces. With the introduction of a warp factor, the two intervals are combined to construct an $SO(2,D-1)$ invariant interval. The embedding method is then employed to obtain the $SO(2,D)$ invariant action for the $AdS_d$ probe brane.

The background vielbein and spin connection of the $AdS_d$ world volume can be found by considering the Maurer-Cartan one form made from just the (pseudo-) translation 
$SO(2,p)/SO(1,p)$ coset element 
\be
\bar\Omega = e^{ix^m P_m} .
\ee
The $AdS_d$ covariant coordinate differential, $\bar\omega^m$, and spin connection, $\bar\omega_M^{mn}$ are obtained from the $AdS_d$ coordinate one-form
\bea
\bar\omega &=& \bar\Omega^{-1} d \bar\Omega \cr
 &=&i\left[ \bar\omega^m P_m +\bar\omega_{M}^{mn} M_{mn} \right] ,
\eea
where
\bea
\bar\omega^m &=& \frac{\sin{\sqrt{m^2x^2}}}{\sqrt{m^2x^2}} P_{T}^{mn}(x) dx_n  +P_{L}^{mn}(x) dx_n \cr
\bar\omega_M^{mn} &=& \left[ \cos{\sqrt{m^2x^2}} -1\right] \frac{(x^m dx^n -x^n dx^m)}{x^2},
\label{AdSdoneform}
\eea
with the transverse and longitudinal projectors for $x^\mu$ defined as in equation (\ref{xprojectors}).
The differential $\bar\omega^m$ is related to the $x^\mu$ world volume coordinate differential via the $AdS_d$ background vielbein $\bar{e}_\mu^{~m}(x)$ as
\be
\bar\omega^m = dx^\mu \bar{e}_\mu^{~m}(x) .
\ee  
Using equation (\ref{AdSdoneform}) along with $d=dx^\mu \partial_\mu^x$, the vielbein presented in equation (\ref{AdS_dvielbein}) is obtained.

Likewise, the vielbein and hence metric for the covolume Euclidean $AdS_N$ space can be obtained by considering the Maurer-Cartan one form made from just the broken (pseudo-) translation 
$SO(1,N)/SO(N)$ coset element, $\phi^i$ being the covolume coordinates,
\be
\bar{\bar\Omega} = e^{i\phi^i Z_i} .
\ee
The covolume covariant coordinate differential, $\bar\omega^i$, is obtained from the one-form
\be
\bar{\bar\omega}= \bar{\bar\Omega}^{-1} d\bar{\bar\Omega} =i\left[ \bar\omega^i Z_i + \frac{1}{2}\bar\omega_T^{ij} T_{ij}\right] ,
\ee
where
\bea
\bar\omega^i &=& d\phi^j \bar{e}_j^{~i} \cr
\bar\omega_T^{ij} &=& \left[ \cosh{\sqrt{m^2\phi^2}} -1\right] \frac{(\phi^i d\phi^j -\phi^j d\phi^i)}{\phi^2} ,
\eea
with the covolume vielbein $\bar{e}_j^{~i}$ given in equation (\ref{AdS_Nvielbein}).

In order to obtain the $SO(2,D-1)$ invariant action for the probe $p$-brane most directly, consider the $SO(2,D-1)$ invariant interval in $D$ dimensional $AdS$ space with coordinates $X^{\cal M} = (x^\mu , \phi^i)$ 
\be
ds^2 = dX^{\cal M} \bar{g}_{\cal MN} dX^{\cal N} =  e^{2A(\phi)}d\bar{s}^2 - dr^2,
\label{interval4}
\ee
where the warp factor $A(\phi)= \ln{\cosh{\sqrt{(m^2\phi^2)}}}$ is introduced and $d\bar{s}^2 = dx^\mu \bar{g}_{\mu\nu}(x) dx^\nu$ is the $AdS_d$ invariant interval with $\bar{g}_{\mu\nu}$ the $AdS_d$ world volume metric tensor, $\bar{g}_{\mu\nu} = \bar{e}_\mu^{~m} \eta_{mn} \bar{e}_\nu^{~n}$.   Similarly, $dr^2 = d\phi^i \bar{g}_{ij}d\phi^j $ is the invariant interval for the Euclidean $AdS_N$ covolume with metric given by $\bar{g}_{ij}= \bar{e}_{i}^{~k}\delta_{kl} \bar{e}_{j}^{~l}$.  

The static gauge coordinate motions of the $p$-brane into the covolume are identified with $N$ Nambu-Goldstone fields $\phi^i (x)$. 
The induced metric on the probe brane's world volume, $g_{\mu\nu}$, with $SO(2,D-1)$ isometries follows from
\be
ds^2 =dx^\mu g_{\mu\nu}dx^\nu = dx^\mu \left[ e^{2A(\phi)}\bar{g}_{\mu\nu} - \partial_\mu \phi^i \bar{g}_{ij}\partial_\nu \phi^j\right]dx^\nu ,
\label{inducedg}
\ee
and the $SO(2,D-1)$ invariant action is therefore given by
\be
\Gamma = -\sigma \int d^d x \sqrt{(-1)^p\det{g}} ~~. \label{invaraction}
\ee
The induced metric, equation (\ref{inducedg}), can be factorized into the product of warped background vielbeine, $e^A\bar{e}_\mu^{~m}$,  and the Nambu-Goto metric, $n_{mn}$, for the Nambu-Goldstone fields
\be
g_{\mu\nu} = \left(e^{A(\phi)}\bar{e}_\mu^{~m}\right) n_{mn} \left(e^{A(\phi)}\bar{e}_{\nu}^{~n}\right) ,
\ee
where
\be
n_{mn}= \eta_{mn} -{\cal D}_m \phi^i (x) \delta_{ij} {\cal D}_n \phi^j (x) ,
\ee
with the  partially covariant derivative ${\cal D}_m \phi^i ={\cal D}_m^{ij} \phi^j$ defined as in equation (\ref{PartCovDer}).
With these definitions, the $SO(2,D-1)$ invariant action (\ref{invaraction}) becomes
\be
\Gamma  = -\sigma \int d^d x \det{\bar{e}}~e^{dA(\phi)}~\sqrt{(-1)^p\det{n}}  .
\ee
The determinant of the Nambu-Goto metric in turn can be expressed in terms of the determinant of a $SO(N)$ matrix
\be
\det{n} =(-1)^p \det{\left[\delta^{ij} -  {\cal D}_m \phi^i \eta^{mn} {\cal D}_n \phi^j \right]} .
\ee
This yields the $AdS_d$ probe $p$-brane $SO(2,D-1)$ invariant action in the form
\be
\Gamma =-\sigma \int d^d x \det{\bar{e}}~\cosh^d{\sqrt{(m^2 \phi^2)}}\sqrt{\det{\left[\delta^{ij} - {\cal D}_m \phi^i \eta^{mn} {\cal D}_n \phi^j  \right]}} ,
\label{NGAction4}
\ee
as obtained in equations (\ref{action1a}) and (\ref{action2a}).

\newsection{\large A Non-BPS Vortex In $D=6$, ${\cal N}=(1, 0)$ Superspace}

The formation of a non-BPS vortex in six dimensional superspace completely breaks the supersymmetry as well as two of the space translation symmetries. 
Such a non-BPS vortex appears, for instance, in 
a supersymmetric gauge theory with adjoint Higgs fields (hypermultiplets) 
(see e.g. \cite{Markov:2004mj}). 
The oscillations of the vortex into superspace is described by two Nambu-Goldstone bosons and four Goldstinos.  
The case of a BPS vortex with partial supersymmetry breaking in the Abelian Higgs model was discussed in \cite{Hughes:1986dn}, while coset methods were used to construct effective actions describing a non-BPS $p=2$ brane embedded in $D=4$, $N=1$ superspace in \cite{Clark:2002bh} and a non-BPS $p=3$ brane embedded in $D=5$, $N=1$ superpace in \cite{Clark:2004jn}. 
In general, $D=6$, ${\cal N}= (1,1)$  superspace has eight supersymmetries with the associated eight component complex (Dirac) spinor supersymmetry charges ${\cal Q}_a~;~a,b = 1 \dots 8$ and $\bar{\cal Q} \equiv {\cal Q}^{\dagger} \Gamma^0$.  The non-vanishing (anti-) commutation relations are
\bea
\left\{ {\cal Q}_a, \bar{\cal Q}^b \right\} & = & 2 \Gamma_b^{{\cal M}b} P_{\cal M} \nonumber \\
\left[ M^{\cal MN}, {\cal Q}_a \right] & = & -\frac{1}{2} \Gamma_a^{{\cal MN}b} {\cal Q}_b \nonumber \\
\left[ M^{\cal MN}, \bar{\cal Q}^b \right] & = & \frac{1}{2} \bar{\cal Q}^a \Gamma_a^{{\cal MN}b} .
\eea
The number of complex supersymmetry generators can be reduced from eight to four using left and right Weyl projectors  defined as
\begin{eqnarray}
{\cal P}_L & \equiv & \frac{1}{2} (1-\Gamma_*) \nonumber \\
{\cal P}_R & \equiv & \frac{1}{2} (1+\Gamma_*),
\end{eqnarray}
with $\Gamma_* = -\Gamma^0 \Gamma^1 \Gamma^2 \Gamma^3 \Gamma^4 \Gamma^5$.  The resulting $(1,0)$ $D=6$ supersymmetry algebra is obtained by constraining the supersymmetry generators to satisfy
${\cal Q}_L ={\cal P}_L {\cal Q}$ and hence ${\cal P}_R {\cal Q}_L =0$ so that the supersymmetry charges obey the algebra
\bea
\left\{ {\cal Q}_{La}, \bar{\cal Q}_L^b \right\} & = & 2 {\cal P}_{La}^{~~~c} \Gamma_c^{{\cal M}b} P_{\cal M} \nonumber \\
\left[ M^{\cal MN}, {\cal Q}_{La} \right] & = & -\frac{1}{2} \Gamma_a^{{\cal MN}b} {\cal Q}_{Lb} \nonumber \\
\left[ M^{\cal MN}, \bar{\cal Q}_L^b \right] & = & \frac{1}{2} \bar{\cal Q}_L^a \Gamma_a^{{\cal MN}b} ~,
\eea
while remaining non-trivial commutators form the Poincar\'e algebra
\bea
\left[M^{\cal MN} , M^{\cal RS} \right] &=& -i \left( \eta^{\cal MR} M^{\cal NS} -\eta^{\cal MS} M^{\cal NR} +\eta^{\cal NS} M^{\cal MR} -\eta^{\cal NR} M^{\cal MS} \right) \cr
\left[M^{\cal MN} , P^{\cal L} \right] &=& i \left( P^{\cal M} \eta^{\cal NL} - P^{\cal N} \eta^{\cal ML} \right) .
\eea

With the formation of the non-BPS vortex \cite{Markov:2004mj} in this $D=6, ~{\cal N}=(1,0)$ superspace, the resulting world volume is just $d=4$ Minkowski space with a bosonic codimension N=2. Thus the 
isometry group of the target superspace is broken by the vortex to the Poincar\'e group of $d=4$ space-time which is $ISO(1,3)$.  As such, all
symmetry charges can be expressed in terms of their $SO(1,3)$ Lorentz group and $SO(2)$ covolume rotation group representation content.  The bosonic charges are given as in the previous sections now with $i=1,2$ for the $SO(2)$ covolume rotation indices so that the $T^{ij} = \frac{1}{2}\epsilon^{ij} T$ obey the algebra of equations (\ref{SO2DAlgebra})-(\ref{Algebra3}) with $m^2 =0$.  The supersymmetry charges form two complex $d=4$ Weyl spinors denoted by $Q_\alpha$, $S_\alpha$ and their complex conjugates $\bar{Q}_{\dot\alpha}$ and $\bar{S}_{\dot\alpha}$.  The left handed Weyl projection of the $D=6$ spinor of supersymmetry charges, ${\cal Q}_{La}$, contains $Q_\alpha$ and $\bar{S}^{\dot\alpha}$, while the conjugate spinor $\bar{\cal Q}_L^a$ contains the charges $S^\alpha$ and $\bar{Q}_{\dot\alpha}$.  The $D=6$ supersymmetry algebra is then given by the centrally extended $d=4$, $N=2$ SUSY algebra with non-vanishing commutators
\bea
\{Q_\alpha , \bar{Q}_{\dot\alpha} \} &=& 2\sigma^\mu_{\alpha\dot\alpha} P_\mu \cr
\{S_\alpha , \bar{S}_{\dot\alpha} \} &=& 2\sigma^\mu_{\alpha\dot\alpha} P_\mu \cr
\{Q_\alpha , {S}^{\beta} \} &=&  -2iZ \delta_\alpha^{~\beta}\cr
\{\bar{Q}_{\dot\alpha} , \bar{S}^{\dot\beta} \} &=&  2i\bar{Z} \delta_{\dot\alpha}^{~\dot\beta}\cr
[M^{\mu\nu} ,Q_\alpha] &=& -\frac{1}{2} \sigma^{\mu\nu \beta}_\alpha Q_\beta \cr
[M^{\mu\nu} ,S_\alpha] &=& -\frac{1}{2} \sigma^{\mu\nu \beta}_\alpha S_\beta \cr
[M^{\mu\nu} ,\bar{Q}^{\dot\alpha}] &=&  \bar\sigma^{\mu\nu \dot\alpha}_{~~~\dot\beta} \bar{Q}^{\dot\beta} \cr
[M^{\mu\nu} ,\bar{S}^{\dot\alpha}] &=&  \bar\sigma^{\mu\nu \dot\alpha}_{~~~\dot\beta} \bar{S}^{\dot\beta} \cr
[T, Q_\alpha ] &=& - Q_\alpha \cr
[T, S_\alpha ] &=& - S_\alpha \cr
[T, \bar{Q}_{\dot\alpha} ] &=&  \bar{Q}_{\dot\alpha} \cr
[T, \bar{S}_{\dot\alpha} ] &=&  \bar{S}_{\dot\alpha} \cr
[\bar{K}^\mu , Q_\alpha] &=& -2\sigma^\mu_{\alpha\dot\alpha} \bar{S}^{\dot\alpha} \cr
[{K}^\mu , \bar{Q}^{\dot\alpha}] &=& 2\bar\sigma^{\mu\dot\alpha\alpha} {S}_{\alpha} \cr
[\bar{K}^\mu , S_\alpha] &=& -2\sigma^\mu_{\alpha\dot\alpha} \bar{Q}^{\dot\alpha} \cr
[{K}^\mu , \bar{S}^{\dot\alpha}] &=& 2\bar\sigma^{\mu\dot\alpha\alpha} {Q}_{\alpha} ~,
\eea
where we have defined $Z=Z^1-iZ^2~, ~\bar{Z}=Z^1+iZ^2~,~K^\mu=K^{1\mu}-iK^{2\mu}~,~\bar{K}^\mu=K^{1\mu}+iK^{2\mu}$.

Denoting the $D=6, ~{\cal N}=(1,0)$ super-Poincar{\'e} group by $G$ and the $d=4$ Poincar{\'e} stability group $ISO(1,3)$ by $H$, all symmetry
transformations can be realized by group elements in $G$ acting on the  $G/H$ coset
element $\Omega$ formed from the 
$P^\mu ,~Z_i ,~K^\mu_i ,~Q_\alpha ,~\bar{Q}^{\dot\alpha} ,~S_\alpha ,~\bar{S}^{\dot\alpha}$ charges as
\be
\Omega (x) \equiv e^{i x^\mu P_\mu } e^{i \phi^i(x) Z_i} e^{i[\theta (x)^\alpha Q_\alpha +\bar\theta_{\dot\alpha} (x) \bar{Q}^{\dot\alpha} +\lambda (x)^\alpha S_\alpha +\bar\lambda_{\dot\alpha} (x) \bar{S}^{\dot\alpha}]} e^{iv^\mu_i(x) K_\mu^i} ,
\ee
where $x^\mu$ are the four dimensional Minkowski space world volume coordinates, 
$\phi^i (x)$ and $v^\mu_i (x)$ are the collective coordinate Nambu-Goldstone bosons associated with the broken $D$ dimensional Poincar\'e symmetries and $\theta_\alpha (x) ,~ \bar\theta^{\dot\alpha}(x) ,~\lambda_\alpha (x) ,~\bar\lambda^{\dot\alpha} (x)$ are the Goldstinos associated with the broken supersymmetries. Collectively, these fields describe the motion of the vortex into the superspace covolume. Using the Maurer-Cartan one-forms, the covariant coordinate differentials and the 
covariant derivatives of all of the Nambu-Goldstone fields can be extracted.  Expanding the Maurer-Cartan one-form in terms of the generators as
\bea
\omega(x) &=& \Omega^{-1} d \Omega = i\left[ \omega^m P_m + \omega_{Zi} Z^i +\omega^\alpha_Q Q_\alpha + \omega_{\bar{Q}\dot\alpha} \bar{Q}^{\dot\alpha} +\omega^\alpha_S S_\alpha + \omega_{\bar{S}\dot\alpha} \bar{S}^{\dot\alpha} \right. \cr
 & &\left. \qquad\qquad\qquad\qquad +\omega^m_{Ki} K_{im} +\frac{1}{2}\omega_M^{mn} M_{mn}+\frac{1}{2}\omega_{T}^{ij} T_{ij}\right],
\eea
yields the vierbein and the covariant derivative of the Nambu-Goldstone bosons $\phi^i$ as 
\bea
\omega^m &=& dx^\mu e_\mu^{~m} \cr
 &=& dX^n N_n^{~m} \cr
\omega_{Zi} &=& dx^\mu \nabla_\mu \phi^i\cr
&=&dX^m \left[ {\nabla}_m \Phi^j (\cosh{\sqrt{4U}}_{ji}-v_m^j (U^{-1/2}\sinh{\sqrt{4U}})_{ji}\right] .
\eea
Here $dX^m = dx^\mu \hat{e}_\mu^{~m}$ is the Akulov-Volkov coordinate differential with $\hat{e}_\mu^{~m}$ the Akulov-Volkov vierbein
\be
\hat{e}_\mu^{~m} = \delta_\mu^{~m} -i\left[ \theta \sigma^m \stackrel{\leftrightarrow}{\partial}_\mu \bar\theta +  \lambda \sigma^m \stackrel{\leftrightarrow}{\partial}_\mu \bar\lambda \right]
\label{AVV}
\ee
and ${\cal D}_m = \hat{e}_m^{-1 \mu} \partial_\mu$ the SUSY covariant derivative.  The Nambu-Goto vierbein $N_n^{~m}$ is obtained as
\bea
N_n^{~m} &=& \delta_n^{~m} + v_m^i \left[ U^{-1/2} (\cosh{\sqrt{4U}}-1)U^{-1/2}\right]_{ij} v^{jm}\cr
 & & \qquad\qquad - {\nabla}_m \Phi^i \left[U^{-1/2} \sinh{\sqrt{4U}}\right]_{ij} v^{jm} ,
\eea
where the Nambu-Goldstone boson differentials $d\Phi^i = dX^m \nabla_m \Phi^i$ are \footnote{The analogous expression in the case of a codimension 1, p=3 brane embedded in $D=5,~ {\cal N}=1$ superspace given in reference \cite{Clark:2004jn} should be amended to a similar form replacing $\partial_\mu( \theta\lambda  -\bar\theta \bar\lambda)$ with $(\theta \stackrel{\leftrightarrow}{\partial}_\mu \lambda-\bar\theta \stackrel{\leftrightarrow}{\partial}_\mu\bar\lambda)$.}
\bea
d\Phi^1 &=& d\phi^1 + \left[(\lambda d\theta -\theta d\lambda) + (\bar\lambda d \bar\theta - \bar\theta d \bar\lambda)\right] \cr
d\Phi^2 &=& d\phi^2 -i\left[(\lambda d\theta -\theta d\lambda) - (\bar\lambda d \bar\theta - \bar\theta d \bar\lambda)\right] .
\eea

The invariant action describing the motion of the 3-brane vortex into the superspace covolume is given by the invariant synthesis of the Akulov-Volkov and Nambu-Goto actions as
\be
\Gamma =-\sigma \int d^4 x \det{e} = -\sigma \int d^4 x \det{\hat{e}} \det{N} ,
\ee
with $\sigma$ the brane tension. Since the action is independent of $v_\mu^i$ derivatives, this field is nondynamical and it can be eliminated using its field equation, or equivalently the inverse Higgs mechanism. So doing yields the covariant constraint $\omega_{Zi}=0$ which allows $v_\mu^i$ to be eliminated in terms of ${\nabla}_m \Phi^i$ as
\be
{\nabla}_m \Phi^i = \left[ U^{-1/2} \tanh{\sqrt{4U}}\right]_{ij}v_m^j .
\ee
Substituting this into the Nambu-Goto vierbein eliminates $v_\mu^i$ from $N_n^{~m}$ giving
\be
N_n^{~m} = \delta_n^{~m} + {\nabla}_n \Phi^i G_{ij} {\nabla}^m \Phi^j ,
\ee
where now the $SO(2)$ matrix $G_{ij}$ (recall equation (\ref{G})) is given in terms of the SUSY covariant derivatives of $\Phi$ as
\be
G_{ij} = H^{-1/2}_{ik}\left[\sqrt{(1-H)}-1\right]_{kl} H^{-1/2}_{lj} ,
\ee
with $H_{ij}$ (recall equation (\ref{H})) expressed as
\be
H_{ij}\equiv {\nabla}_m \Phi_i {\nabla}^m \Phi_j .
\ee
Hence the determinant of the Nambu-Goto vierbein becomes
\be
\det{N} = \det{\sqrt{(1 - H)}} = \det{\sqrt{(\delta_{ij} - {\nabla}^m \Phi_i \eta_{mn} {\nabla}^n \Phi_j )}} ~,
\ee
which after exploiting the $2\times 2$ nature of the $H$ matrices takes the form
\be
\det{N}=\sqrt{\left(1 - {\rm Tr}{[H]}+\det{H} \right)}= \sqrt{\left(1 - {\rm Tr}{[{\nabla}^m \Phi_i {\nabla}_m \Phi_j]}+\det{{\nabla}^m \Phi_i {\nabla}_m \Phi_j} \right)} .
\ee

Thus the final form of the Akulov-Volkov-Nambu-Goto 
invariant action describing the covolume oscillations of a non-BPS vortex embedded in $D=6,~{\cal N}=(1,0)$ superspace is secured as
\bea
\Gamma &=& -\sigma \int d^4 x \det{\hat{e}}\det{\sqrt{(\delta_{ij} - {\nabla}^m \Phi_i \eta_{mn} {\nabla}^n \Phi_j )}} \cr
 &=& -\sigma \int d^4 x \det{\hat{e}} \sqrt{\left(1 - {\rm Tr}{[{\nabla}^m \Phi_i {\nabla}_m \Phi_j]}+\det{{\nabla}^m \Phi_i {\nabla}_m \Phi_j} \right)} \cr
 &=&-\sigma \int d^4 x \det{\hat{e}} \sqrt{\left(1 -{\nabla}^m \Phi_i {\nabla}_m \Phi_i +\frac{1}{2}\epsilon_{ij}\epsilon_{kl}{{\nabla}^m \Phi_i {\nabla}_m \Phi_k{\nabla}^n \Phi_j {\nabla}_n \Phi_l} \right)}  .\cr
 & & 
\label{NGaction8}
\eea
The Akulov-Volkov vierbein $\hat{e}_\mu^{~m}$ is given by equation (\ref{AVV}) and the SUSY covariant derivatives of the Nambu-Goldstone boson fields are given by the differentials $d\Phi^i = dX^m {\nabla}_m \Phi^i$ where
\bea
{\nabla}_m \Phi^1 &=& {\cal D}_m \phi^1 + \lambda \stackrel{\leftrightarrow}{{\cal D}}_m\theta +\bar\lambda \stackrel{\leftrightarrow}{{\cal D}}_m  \bar\theta  \cr
{\nabla}_m \Phi^2 &=& {\cal D}_m \phi^2 -i \lambda \stackrel{\leftrightarrow}{{\cal D}}_m\theta +i\bar\lambda \stackrel{\leftrightarrow}{{\cal D}}_m  \bar\theta  .
\eea

The Nambu-Goldstone bosons can equivalently be described by tensor gauge fields \cite{Hayashi:1973nf,Freedman:1980us,Bagger:1997pi,Clark:2004jn}.  The dual action which involves the replacement of one or both of the scalar fields by tensor fields can be constructed using Lagrange multipliers.  For the mixed scalar-tensor action, a Lagrange multiplier denoted by $L^m$ is introduced and used to define the vector $l_m = \nabla_m\Phi^2$ so that the action (\ref{NGaction8}) (with $\Phi^1$ relabeled as $\varphi$) can then be written as
\be
\Gamma = -\sigma \int d^4 x \det{\hat{e}} \left\{ \sqrt{1-\nabla^\mu \varphi \nabla_\mu \varphi -l^2 + l^2(\nabla\varphi)^2 - (l^\mu \nabla_\mu \varphi)^2} + L^\mu( l_\mu - \nabla_\mu \Phi^2)\right\}.
\ee
The $\Phi^2$ equation of motion yields the identity
\be
\partial_\mu \left( \det{\hat{e}} ~ \hat{e}_m^{-1\mu} L^m \right) =0 
\ee
whose solution is
\be
F^\mu = \det{\hat{e}} ~ \hat{e}_m^{-1\mu} L^m = \epsilon^{\mu\nu\rho\sigma} \partial_\nu B_{\rho\sigma} ~,
\ee
where $B_{\mu\nu}$ is the dual tensor gauge field.  Recalling that 
\be
{\nabla}_m \Phi^2 = \hat{e}_m^{-1\mu} \left\{\partial_\mu \phi^2 -i \lambda \stackrel{\leftrightarrow}{{\partial}}_\mu\theta +i\bar\lambda \stackrel{\leftrightarrow}{{\partial}}_\mu  \bar\theta \right\},
\ee
$\Phi^2$ can be eliminated from the action by an integration by parts to yield
\bea
\Gamma &=& -\sigma \int d^4 x \det{\hat{e}} \left\{ \sqrt{1-\nabla^\mu \varphi \nabla_\mu \varphi -l^2 + l^2(\nabla\varphi)^2 - (l^\mu \nabla_\mu \varphi)^2} \right.\cr
 & &\left.\qquad\qquad\qquad\qquad + L^\mu l_\mu +iL^m \hat{e}_m^{-1\mu}(\lambda \stackrel{\leftrightarrow}{{\partial}}_\mu\theta -\bar\lambda \stackrel{\leftrightarrow}{{\partial}}_\mu  \bar\theta) \right\}.
\eea
The $l^\mu$ equation of motion
\be
L^a = \frac{\left[ l^m(1-(\nabla\varphi)^2 + (l\cdot\nabla\varphi) \nabla^m \varphi\right]}{\sqrt{(1-(\nabla \varphi)^2 -l^2 + l^2(\nabla\varphi)^2 - (l^\mu \nabla_\mu \varphi)^2)}},
\ee
can then be used to eliminate the $l^\mu$ field to obtain the mixed scalar field $\varphi$ and tensor field $L^m = (1/\det{\hat{e}})\hat{e}_\mu^{~m} \epsilon^{\mu\nu\rho\sigma} \partial_\nu B_{\rho\sigma}$ dual action 
\cite{Bagger:1997pi}
\bea
\Gamma &=& -\sigma \int d^4 x \det{\hat{e}} \left\{ \sqrt{1-\nabla^\mu \varphi \nabla_\mu \varphi +L^m L_m + (L^m \nabla_m\varphi)^2 } \right.\cr
 & &\left.\qquad\qquad\qquad\qquad\qquad\qquad +iL^m \hat{e}_m^{-1\mu}(\lambda \stackrel{\leftrightarrow}{{\partial}}_\mu\theta -\bar\lambda \stackrel{\leftrightarrow}{{\partial}}_\mu  \bar\theta) \right\}.
\label{action9}
\eea

Finally, the completely dual tensor gauge field action can be obtained by introducing another Lagrange multiplier $K^m$ to define the vector field $k_m = \nabla_m \varphi = \hat{e}_m^{-1\mu}[\partial_\mu \phi^1 +(\lambda \stackrel{\leftrightarrow}{{\partial}}_\mu\theta +\bar\lambda \stackrel{\leftrightarrow}{{\partial}}_\mu  \bar\theta)]$.  With these substitutions the action (\ref{action9}) can be written as
\bea
\Gamma &=& -\sigma \int d^4 x \det{\hat{e}} \left\{ \sqrt{1-k^m k_m +L^m L_m - (L^m k_m)^2 } \right.\cr
 & &\left.\qquad +iL^m \hat{e}_m^{-1\mu}(\lambda \stackrel{\leftrightarrow}{{\partial}}_\mu\theta -\bar\lambda \stackrel{\leftrightarrow}{{\partial}}_\mu  \bar\theta)+K^m(k_m - \nabla_m\varphi) \right\}.
\eea
As previously, the $\varphi$ field equation $\partial_\mu [\det{\hat{e}}~\hat{e}_m^{-1\mu} K^m]=0$ is solved in terms of the second tensor gauge field giving
\be
G^\mu = \det{\hat{e}}~\hat{e}_m^{-1\mu} K^m = \epsilon^{\mu\nu\rho\sigma} \partial_\nu C_{\rho\sigma} .
\ee
The scalar field $\varphi$ can then be eliminated using integration by parts to obtain the action
\bea
\Gamma &=& -\sigma \int d^4 x \det{\hat{e}} \left\{ \sqrt{1-k^m k_m +L^m L_m - (L^m k_m)^2 }+K^m k_m \right.\cr
 & &\left.\qquad +iL^m \hat{e}_m^{-1\mu}(\lambda \stackrel{\leftrightarrow}{{\partial}}_\mu\theta -\bar\lambda \stackrel{\leftrightarrow}{{\partial}}_\mu  \bar\theta) - K^m \hat{e}_m^{-1\mu}(\lambda \stackrel{\leftrightarrow}{{\partial}}_\mu\theta +\bar\lambda \stackrel{\leftrightarrow}{{\partial}}_\mu  \bar\theta) \right\}.
\eea
Lastly, use of the $k_m$ equation of motion 
\be
K^m = \frac{k^m +(L^n k_n) L^m}{\sqrt{1-k^m k_m +L^m L_m - (L^m k_m)^2 }}
\ee
allows the complete dual form of the action to be secured as
\bea
\Gamma &=& -\sigma \int d^4 x \det{\hat{e}} \left\{ \sqrt{(1+L^2)(1+K^2)- (L\cdot K)^2} \right.\cr
 & &\left.\qquad +iL^m \hat{e}_m^{-1\mu}(\lambda \stackrel{\leftrightarrow}{{\partial}}_\mu\theta -\bar\lambda \stackrel{\leftrightarrow}{{\partial}}_\mu  \bar\theta) - K^m \hat{e}_m^{-1\mu}(\lambda \stackrel{\leftrightarrow}{{\partial}}_\mu\theta +\bar\lambda \stackrel{\leftrightarrow}{{\partial}}_\mu  \bar\theta) \right\}.
\eea

The $SO(2)$ invariance of the action can be made more manifest by introducing the notation
\bea
V_1^m &=& K^m \cr
V_2^m &=& L^m \cr
J_{1m}&=& \hat{e}_m^{-1\mu} (\lambda \stackrel{\leftrightarrow}{{\partial}}_\mu\theta +\bar\lambda \stackrel{\leftrightarrow}{{\partial}}_\mu  \bar\theta) \cr
J_{2m}&=& \hat{e}_m^{-1\mu}(\lambda \stackrel{\leftrightarrow}{{\partial}}_\mu\theta -\bar\lambda \stackrel{\leftrightarrow}{{\partial}}_\mu  \bar\theta) 
\eea
in terms of which the action reads
\bea
\Gamma &=& -\sigma \int d^4 x \det{\hat{e}} \left\{ \det{\sqrt{(1 + V^m_i V_{jm})}} -V^m_i J_{mi}\right\} \cr
 &=&-\sigma \int d^4 x \det{\hat{e}} \left\{ \sqrt{1 + V^m_i V_{im} + \frac{1}{2}\epsilon_{ij}\epsilon_{kl} (V^m_i V_{km})(V^n_j V_{ln})} -V^m_i J_{mi}\right\} .
\eea
Using the Akulov-Volkov vierbein to convert tangent space indices to world volume indices and visa versa so that, for example, $ J^\mu_i = \hat{e}_m^{-1\mu} J_i^m$, and introducing the Akulov-Volkov metric $\hat{g}_{\mu\nu} = \hat{e}_\mu^{~m} \eta_{mn} \hat{e}_\nu^{~n}$ and the world volume tensor density fields $F^\mu_i = \det{\hat{e}}~ \hat{e}_m^{-1\mu} V^m_i$ so that $F_{i\mu} = (1/\det{\hat{e}}) \hat{g}_{\mu\nu} F^\nu_i$, the action takes the final form
\bea
\Gamma &=& -\sigma \int d^4 x \left\{ \sqrt{-\det{\hat{g}}}~\sqrt{1+ F_{i\mu} \hat{g}^{\mu\nu} F_{i\nu} + \det{(F_{i\mu} \hat{g}^{\mu\nu} F_{j\nu})}} - F_{i\mu} J^\mu_i \right\} \cr
 &=&-\sigma \int d^4 x \left\{ \sqrt{-\det{\hat{g}}}~\sqrt{\det{(1+ F_{i\mu} \hat{g}^{\mu\nu} F_{j\nu})}} - F_{i\mu} J^\mu_i \right\} \cr
 &=& -\sigma \int d^4 x \left\{ \sqrt{-\det{[\hat{g}_{\mu\nu}+ F_{i\mu} F_{i\nu}]}} -F_{i\mu} J^\mu_i\right\} .
\eea

If, in addition to translation symmetries, the presence of the vortex leads to the spontaneous breakdown of various global internal symmetries, then the
effective world volume theory also contains the associated
Nambu-Goldstone scalar fields, while the dual theory is a function of the corresponding
non-Abelian tensor gauge fields \cite{Freedman:1980us}.
A vortex of this type occurs in supersymmetric 
$U(N_C)$ gauge theory with $N_F=N_C$ fundamental hypermultiplets 
in $D=4$ \cite{Hanany:2003hp} 
and also in its $D=6$ \cite{Eto:2004ii} 
dimensional generalization.

\vspace*{1.0in}
\noindent The work of TEC, STL and CX was supported in part by the U.S. Department of Energy under grant DE-FG02-91ER40681 (Task B).  The work of TtV was supported in part by a Cottrell Award from the Research Corporation.  TtV would like to thank the theoretical physics group at Purdue University for their hospitality during his sabbatical leave from Macalester College.

\newpage
\end{document}